\documentclass[12pt]{article}
\usepackage{epsf}
\usepackage{amsmath}
\usepackage{graphics}
\usepackage{cite}

\renewcommand{\thefootnote}{\#\arabic{footnote}}
\begin{document}

\newcommand{\gtrsim}{ \mathop{}_{\textstyle \sim}^{\textstyle >} }
\newcommand{\lesssim}{ \mathop{}_{\textstyle \sim}^{\textstyle <} }

\newcommand{\rem}[1]{{\bf #1}}

\renewcommand{\thefootnote}{\fnsymbol{footnote}}
\setcounter{footnote}{0}
\begin{titlepage}

\def\thefootnote{\fnsymbol{footnote}}

\begin{center}
\hfill December 2013

\vskip .5in
\bigskip
\bigskip
{\Large \bf Quiver Approach to Massive Gauge Bosons
Beyond the Standard Model}

\vskip .45in

{\bf Paul Howard Frampton}\footnote{paul.h.frampton@gmail.com}

\bigskip
\bigskip

{Department of Physics and Astronomy, University of North Carolina,
Chapel Hill, NC 27599-3255.}

\end{center}

\vskip .4in

\begin{abstract}
We address the question of the possible existence
of massive gauge bosons beyond
the $W^{\pm}$ and $Z^{0}$ of the standard model.
Our intuitive and aesthetic approach is based on quiver theory.
Examples thereof arise, for example, from compactification of the
type IIB superstring on $AdS_5 \times S_5/ Z_n$
orbifolds. We explore the quiver theory framework more generally
than string theory.
The practical question is what gauge bosons to look
for at the upgraded LHC, in terms of 
color and electric charge, and of their couplings
to quarks and leptons. Axigluons and bileptons are favored.
\end{abstract}

\end{titlepage}

\renewcommand{\thepage}{\arabic{page}}
\setcounter{page}{1}
\renewcommand{\thefootnote}{\#\arabic{footnote}}

\newpage

\section{Introduction}

\bigskip
\bigskip

\noindent
With the discovery\cite{ATLAS,CMS}
of the BEH scalar particle, all the
particles in the standard model (SM) have been found. The
practical question therefore is which particles,
especially massive gauge bosons, are waiting to be
discovered in $pp$-collisions at a center of mass
energy $\sqrt{s} = 14$ TeV?

\bigskip

\noindent
Of the twelve known gauge bosons associated
with the group $G_{SM} = SU(3)_C \times SU(2)_L \times U(1)_Y$
the eight gluons of unbroken color $G_{QCD} = SU(3)_C$ are
massless, as is the electromagnetic photon of
unbroken $U(1)_{em}$ which survives the symmetry
breaking of the electroweak group
$G_{EW} = SU(2)_L \times U(1)_Y \rightarrow U(1)_{em}$.
The remaining three known gauge bosons are massive
with mass arising from the BEH mechanism \cite{BE,H}
so that \cite{PDG} the $W^{\pm}$ has mass $M_W = 80.385 \pm 0.015$ GeV
and $Z^{0}$ has mass $M_Z = 91.1876 \pm 0.0021$ GeV.

\bigskip

\noindent
In the present article, we shall assume that at energy scales
accessible to the LHC the gauge group is a semi-simple group
or more specifically a quasi-simple group $G_{quiver} \ni G_{SM}$
with $G_{quiver} = SU(N)^n$. To accommodate $G_{QCD}$, it
is suggested to identify $N = 3$, whereupon $G_{quiver} = SU(3)^n$.
We are, to some extent, divorcing quiver theory from string theory
by omitting a non-simple factor $U(1)^n$ which is present when
the Type IIB superstring is compactified on the orbifold
$AdS_5 \times S_5/ Z_n$ \cite{PhysicsReports}. For the case
$n = 1$ the $U(1)$ can be rotated away since none of the matter
fields carry a charge under it. For all $n \geq 2$, however,
the $U(1)$'s are present and cause at least two difficulties:
firstly, there are uncanceled triangle anomalies; secondly,
the associated renormalization group (RG) beta functions are
positive definite thus precluding ultra violet (UV) conformality.

\bigskip

\noindent
UV conformality is an underlying motivation but can, at best,
hold good only within some conformality "window" covering 
a finite energy range because
at the Planck energy gravitation enters and necessarily
breaks conformal invariance since Newton's constant is dimensionful.
There is a significant no-go theorem\cite{Klebanov} for
non-SUSY ${\cal N} = 0$ quiver theory showing that within certain
strong assumptions there exist double-trace operators whose
couplings are non-conformal. Nevertheless, this theorem 
makes both a one-loop approximation and a leading-order 1/N
approximation so that conformality remains an open question
when one allows cancelation between different loop orders
and/ or (probably "and") different orders in 1/N.

\bigskip

\noindent
We shall focus primarily on the ${\cal N} = 0$ non-SUSY quiver theories
because the experiments (LHC) show no indication of weak-scale
supersymmetry. ${\cal N} = 2$ theories are generally non-chiral
and phenomenologically disfavored, while ${\cal N} = 1$ theories
can be chiral and more readily possess UV conformality.

\bigskip

\noindent
We shall omit all gravitational effects. The spacetime dimension will be
anchored at four normal bosonic flat dimensions. Our analysis
is deliberately ultraconservative although we are using quiver ideas
which stem in part from speculative directions.

\bigskip

\noindent
We shall study separately the color sector which is simple and
straightforward, restricted to only two possibilities (QCD and
chiral color)
then the much richer electroweak sector.

\bigskip

\section{Color and electroweak sectors}

\noindent
We consider two possibilities for the color part of the quiver. Firstly,
we may take simply $G_{color} = G_{QCD}$ in which case
the remainder of the gauge group is $SU(3)^{(n-1)}$. In this case,
there are no additional gauge bosons beyond the massless eight gluons.
Alternatively, we may assign two quiver nodes to color in the style
of chiral color. In this case, there are eight additional massive
gauge bosons which are a color octer of axigluons. The remainder of the
quiver gauge group is $SU(3)^{(n-2)}$. The axigluons are the only
examples of massive gauge bosons with color that we shall encounter.
Axigluons have no electric charge. The current lower limit on the
axigluon mass from LHC data is at least 3 TeV\cite{LHCaxigluon}.

\bigskip

\noindent
The remainder of the quiver gauge group is $SU(3)^m$
where $m = (n - 1)$ (for QCD) or $m = (n - 2)$ (for
chiral color). We shall normalize the Gell-Mann SU(3)
matrices by Tr$(\lambda_a \lambda_b) = \frac{1}{2} \delta_{ab}$,
so that for the diagonal generators $\lambda_{3, 8}$
to be used in the electric charge $Q$ we shall always mean

\begin{equation}
\lambda_3 = \left( \frac{1}{2} \right)
\left( \begin{array}{ccc} 1 & 0 & 0 \\
0 & -1 & 0 \\
0 & 0 & 0
\end{array}
\right)
\end{equation}

\begin{equation}
\lambda_8 = \left( \frac{1}{2 \sqrt{3}} \right)
\left( \begin{array}{ccc} 1 & 0 & 0 \\
0 & 1 & 0 \\
0 & 0 & -2 
\end{array}
\right)
\end{equation}

\bigskip

\noindent
For the $m$ $SU(3)$ factors we shall label the generators by
$\lambda_a^{(i)}$ with $1 \leq i \leq m$. We may without loss
of generality embed naturally the $SU(2)_L$ of $G_{SM}$ in
the first $SU(3)$ so that $(T_3)_L \equiv \lambda_3^{(1)}$.
We may then rewrite the weak hypercharge $Y$ as

\begin{eqnarray}
Q & = & (T_3)_L + \frac{1}{2} Y \nonumber \\
  & = & \lambda_3^{(1)} +
\Sigma_{i=2}^{i=m} C_3^i \lambda_3^{(i)}
+
\Sigma_{i=1}^{i=m} C_8^i \lambda_8^{(i)}
\label{Q}
\end{eqnarray}

\bigskip

\noindent
To proceed, we shall borrow the quiver rules\cite{Frampton,Vafa,PhysicsReports}
which we  carry over from the orbifolding of the Type IIB superstring
on $AdS_5 \times S_5/ Z_n$. We must specify four integers which show
the embedding of $Z_n$ in the $SU(4)$ which acts on the ${\cal N} = 4$
supersymmetries. We write

\begin{equation}
{\bf 4} = A_{\mu} = (A_1, A_2, A_3, A_4)
\end{equation}
where $\Sigma_{\mu=1}^{\mu=4} A_{\mu} = 0$ (mod n). To 
have a non-SUSY ${\cal N} = 0$
quiver gauge theory all of the $A_{\mu}$ must be non-vanishing (mod n). 

\bigskip

\noindent
From this {\bf 4} of $SU(4)$ we construct the real {\bf 6} = {\bf 3}
 + {\bf 3*}
with

\begin{equation}
{\bf 3} = a_i = (a_1, a_2, a_3)
\end{equation}
where  

\begin{eqnarray}
a_1 & = & A_2 + A_3  \nonumber \\
a_2 & = & A_3 + A_1  \nonumber \\
a_3 & = & A_1 + A_2
\end{eqnarray}

\bigskip

\noindent
With these definitions, the chiral fermions are in the 
representation ${\cal R}_F$:

\begin{equation}
{\cal R}_F = \Sigma_{j=1}^{j=n} \Sigma_{\mu=1}^{\mu = 4} 
(3_j, \bar{3}_{j+A_{\mu}})
\label{RF}
\end{equation}
while the complex scalars  are in the related representation
${\cal R}_S$
\begin{equation}
{\cal R}_S = \Sigma_{j=1}^{j=n} \Sigma_{i=1}^{i=3} (3_j, \bar{3}_{j\pm a_i})
\end{equation}

\bigskip

\noindent
The chiral fermions (oriented) and complex scalars (non-oriented)
are conveniently displayed on a quiver diagram with $n$ nodes.
The representations ${\cal R}_F$ and ${\cal R}_S$ are related
so that the Yukawa couplings correspond to triangles with
two sides being oriented chiral fermions and the third side
being a non-oriented complex scalar.

\bigskip

\noindent
The gauge bosons which are in $SU(3)$ octets
at each node. In particular, we study the electric charges
$Q$ according to Eq.(\ref{Q}) and the couplings to chiral fermions
which enter and leave the node according to their
representation ${\cal R}_F$ given in Eq.(\ref{RF}).

\bigskip

\noindent
If the electric charges of the particles in the defining
{\bf 3} are $Q = (q_1, q_2, q_3)$ then the corresponding
charges of the eight gauge bosons are
$\pm (q_1 - q_2), \pm (q_2 - q_3), \pm (q_3 - q_1),$ together
with two neutrals.
For example, taking into account only the contribution to $Q$
from the first $m = 1$ $SU(3)$ factor and assigning $C_8^1 = \sqrt{3}$
one finds for the first {\bf 3} that $Q=(+1, 0, -1)$
and the gauge boson electric charges are
$++,+,+,0,0,-,-,--$. This is the simplest possibility
which will continue for all subsequent $SU(3)$ s, $i = 2, 3, ....$
as long as $C_3^i=1$ and $C_8^i =\sqrt{3}$.

\bigskip

\noindent
The promotion from $SU(2)_L$ to $SU(3)_L$ will generally, although
not always, lead to
double electric charges for the new massive gauge bosons, so
this is what we can fasten upon in our attempt to make the most
likely predictions for additional particles.

\bigskip

\section{More than three massive gauge bosons}

\bigskip

\noindent
The discovery of the $W^{\pm}$ and $Z^0$ massive gauge bosons
in 1983 provided a watershed which confirmed the correctness
of the SM. Furthermore, the masses
\cite{PDG} $M_W = 80.385 \pm 0.015$ GeV
and $M_Z = 91.1876 \pm 0.0021$ GeV were consistent with the theory
of spontaneous symmetry breaking via the BEH mechanism.

\bigskip

\noindent
One oft speculated additional massive gauge boson is a $Z^{'0}$
arising from an extra $U(1)$ gauge group. But this makes the
gauge group less simple. Aesthetically and intuitively
$G_{SM} = SU(3) \times SU(2) \times U(1)$ has the unsatisfying
feature of not being even semi-simple. UV conformality strongly
disfavors any $U(1)$ gauge factor, in favor of a gauge group
with only non-abelian factors. Even more attractive
is a quasi-simple gauge group like $SU(3)^n$ which with
a discrete symmetry needs only one coupling constant.
Of course, quasi-simple quivers may also contain $Z^{'0}$-type
gauge bosons, but we are more interested in massive gauge bosons
with nontrivial color and electric
charge under the unbroken vacuum gauge symmetry
$G_{VACUUM} \equiv G_{QCD} \times G_{em} \equiv 
SU(3)_C \times U(1)_{em}$.

\bigskip

\noindent
To examine the fermion couplings of the doubly electric charged
$Y^{\pm\pm}$ gauge bosons, consider the case with electroweak
gauge group $SU(3)_L \times U(1)_X$ and

\begin{equation}
Q = \lambda_3^{(1)} + \sqrt{3} \lambda_8^{(1)} + X
\end{equation}

\bigskip

\noindent
The leptons are in $X = 0$ antitriplets of $SU(3)_L$ as

\begin{equation}
\left( \begin{array}{c} e^+ \\ \nu_e \\ e^- 
\end{array}
\right) 
\left( \begin{array}{c} \mu^+ \\ \nu_{\mu} \\ \mu^- 
\end{array}
\right) 
\left( \begin{array}{c} \tau^+ \\ \nu_{\tau} \\ \tau^- 
\end{array}
\right) 
\end{equation}

\bigskip

\noindent
In the quark sector, $Y^{\pm\pm}$ must couple to exotic
quarks with electric charges $- 4/3$ or $+ 5/3$
so the best signature must be purely bileptonic
like $Y^{\pm\pm} \rightarrow \mu^{\pm}\mu^{\pm}$.

\bigskip 
\noindent
Such a quiver theory may be UV conformally invariant, meaning
that when gravity is included there will be a conformality
window as discussed earlier. Such a theory is free of 
triangle anomalies and renormalizable, just like the SM.
Using only the criteria of quiver theory, we expect the simplest
extensions of $G_{SM} = SU(3) \times SU(2) \times U(1)$
to be $SU(3)^3$ with QCD or $SU(3)^4$ with chiral color.

\bigskip

\noindent
In conclusion, assuming that the gauge group $G_{SM}$ fills
out to $SU(N)^n$ where probably $N = 3$ and $n = 3, 4, ...$
then the most likely additional massive gauge bosons
which transform nontrivially under $G_{VACUUM}$
are either (i) color-octet axigluons and/ or (ii) doubly
electrically charged bileptons $Y^{\pm\pm}$ which
could be observed at the upgraded LHC as resonant states 
respectively in (i) dijets and (ii) like-sign lepton pairs.
Present lower limits on their masses are respectively
(i) 3 TeV \cite{LHCaxigluon} and (ii) 850 GeV \cite{willmann}
so higher masses await
investigation and LHC data are eagerly awaited. 

\begin{center}

\section*{Acknowledgement}

\end{center}

\noindent
This work was supported in part
by U.S. Department of Energy Grant DE-FG02-06ER41418.

\end{document}